\documentclass[a4paper,onecolumn,11pt,unpublished]{quantumarticle}
\pdfoutput=1

\usepackage[numbers,sort&compress]{natbib} 
\usepackage[utf8]{inputenc}
\usepackage[english]{babel}
\usepackage[T1]{fontenc}
\usepackage{amsmath}
\usepackage{hyperref}
\usepackage{dcolumn}
\usepackage{bm}
\usepackage{physics}
\usepackage{graphicx}
\graphicspath{{figures/}}

\begin{document}

\title{A perspective on the pathway to a scalable quantum internet using rare-earth ions}

\author{Robert M. Pettit}
\affiliation{memQ Inc., 5235 S. Harper Ct. 8th Floor, Chicago, IL, 60615, USA}
\email{robert.pettit@memq.tech}

\author{Farhang Hadad Farshi}
\affiliation{Da Vinci Labs, 2 c\^{o}te de la Gu\^{e}pi\`{e}re, 37530 Nazelles-Negron, France}

\author{Sean E. Sullivan}
\affiliation{memQ Inc., 5235 S. Harper Ct. 8th Floor, Chicago, IL, 60615, USA}

\author{\'{A}lvaro V\'{e}liz-Osorio}
\affiliation{Da Vinci Labs, 2 c\^{o}te de la Gu\^{e}pi\`{e}re, 37530 Nazelles-Negron, France}

\author{Manish Kumar Singh}
\affiliation{memQ Inc., 5235 S. Harper Ct. 8th Floor, Chicago, IL, 60615, USA}
\email{manish@memq.tech}

\begin{abstract}
The ultimate realization of a global quantum internet will require advances in scalable technologies capable of generating, storing, and manipulating quantum information. The essential devices that will perform these tasks in a quantum network are quantum repeaters, which will enable the long-range distribution of entanglement between distant network nodes. In this perspective, we provide an overview of the primary functions of a quantum repeater and discuss progress that has been made toward the development of repeaters with rare-earth ion doped materials while noting challenges that are being faced as the technologies mature. We give particular attention to erbium, which is well suited for networking applications. Finally, we provide a discussion of near-term benchmarks that can further guide rare-earth ion platforms for impact in near-term quantum networks. 
\end{abstract}

\maketitle

\section{Introduction}
The quantum internet promises to revolutionize computing, communication, and sensing \cite{wehnerQuantumInternetVision2018,simonGlobalQuantumNetwork2017,kimbleQuantumInternet2008}. The processing power offered by \textit{networked} quantum computers could be scaled beyond what is possible with a single quantum computer, finding useful applications in problems of materials science, pharmaceuticals, and optimization among others that classical computers would find intractable. Additionally, a quantum internet would allow communication between users where the security of transmitted data can be guaranteed by the laws of quantum physics \cite{xuSecureQuantumKey2020}. This connectivity could also be exploited to link up measurement devices such as atomic clocks \cite{komarQuantumNetworkClocks2014} or telescopes \cite{gottesmanLongerBaselineTelescopesUsing2012} to advance the state-of-the-art in timekeeping and astronomy, as well as to test the fundamentals of quantum mechanics across unprecedented length scales \cite{rideoutFundamentalQuantumOptics2012}.

A quantum internet will consist of nodes of quantum processors and channels to distribute entanglement between the nodes. Each channel will therefore be required to preserve fragile quantum entanglement. This demand is a steep challenge to meet. The ultimate size of a quantum internet will depend on how well the channels are able to fulfill this task.

\begin{figure}
    \centering
    \includegraphics[scale=1.0]{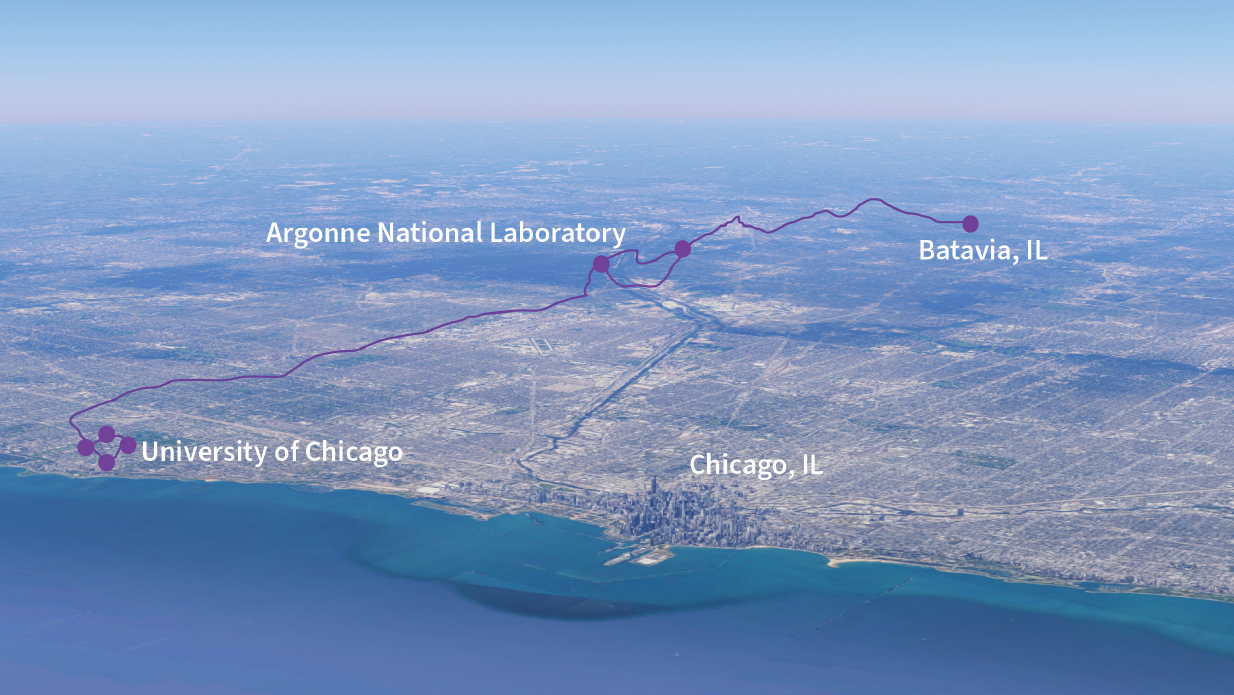}
    \caption{An urban-scale quantum network in Chicago, IL, USA. The network nodes span 200 km of optical fiber. Image courtesy of the Chicago Quantum Exchange.}
    \label{fig:chi_net}
\end{figure}

An optical photon is the ideal mediator of entanglement throughout the network due to its ability to propagate through existing low-loss networks of optical telecommunications fiber. Photons may also be beamed across free-space links \cite{ursinEntanglementbasedQuantumCommunication2007} or to satellites from ground-based nodes \cite{yinSatellitebasedEntanglementDistribution2017}. In this review, we focus on fiber network applications with the view that, in the near-term, optical fiber connections will be used to span local networks with links over 100s of km, while satellites may be reserved to connect more distant nodes separated by 1000s of km or more \cite{awschalomRoadmapQuantumInterconnects2022,awschalomLongdistanceEntanglementBuilding2020}. Construction of urban-scale fiber networks for quantum networking is ongoing in various metropolitan areas around the world to establish test beds for exploring quantum communications technologies under real world operating conditions \cite{peevSECOQCQuantumKey2009,stuckiLongtermPerformanceSwissQuantum2011,sasakiFieldTestQuantum2011,maoIntegratingQuantumKey2018,dynesCambridgeQuantumNetwork2019,awschalomLongdistanceEntanglementBuilding2020}. These networks may then provide the necessary initial infrastructure for building out larger networks to develop a global scale quantum internet. As an example, figure \ref{fig:chi_net} shows an urban-scale network constructed across the Chicago, IL, USA metro area that connects nodes across 200 km of optical fiber.

Even in state-of-the art low-loss fiber, photon losses scale exponentially at rates of about 0.2 dB/km. For wavelengths outside of the telecom window, the loss rates are higher. In classical telecommunications networks, repeaters are used to boost the signal to overcome this inherent loss in the fibers. A full-scale quantum internet will therefore need nodes that can connect via optical fiber, as well as repeaters that can assist with distributing entanglement between the nodes by fighting losses present in the fiber channels. These quantum repeaters will be fundamentally different from repeaters in classical telecommunications networks, however, because the no-cloning theorem prevents the amplification of arbitrary quantum states \cite{woottersSingleQuantumCannot1982}.  Quantum repeaters will have three main essential capabilities: (i) the ability to produce entanglement between stationary matter qubits and photonic “flying” qubits, (ii) the ability to store arbitrary quantum states in memory for asynchronous entanglement creation between distant nodes, and (iii) the ability to perform entanglement swapping operations to distribute entanglement across the network.   

On a physical level, an ideal quantum repeater would deliver on-demand single or entangled photons at a high rate for fast data transfer, operate at telecom wavelengths for long-distance transmission over optical fiber, and exhibit long optical or spin coherence times to enable entanglement operations over timescales commensurate with the time-of-flight of photons across the optical channels. A variety of modalities are under development to enable a quantum internet via optical channels, in particular modalities that utilize atom-like emitters in the solid-state  \cite{aharonovichSolidstateSinglephotonEmitters2016,awschalomQuantumTechnologiesOptically2018}. A major challenge is that all of these modalities face a trade off between the photon emission wavelength, photon emission rate, and the photon or spin coherence times that ultimately limit the suitability of the modality for realizing a quantum repeater. While some modalities, such as semiconductor quantum dots or vacancy complexes in materials such as diamond and SiC, meet one or two of these criteria, few can meet all three without additional engineering. Rare-earth ions embedded in a solid-state host have demonstrated emission at telecom wavelengths as well as long optical and spin coherence times, however they do not inherently deliver photons at a fast rate. The limited photon emission rate can be overcome by engineering on-chip cavities and waveguides to achieve Purcell enhancement \cite{martiniAnomalousSpontaneousEmission1987} and efficient routing of the emission. The combination of rare-earth ions embedded in solid-state hosts with engineered nanophotonic devices can therefore achieve all three necessary criteria to develop quantum repeaters.

Toward this goal, advances in engineering with rare-earth ion platforms have also demonstrated the critical ability to integrate with scalable photonics platforms, either in silicon \cite{przybylinskaOpticallyActiveErbium1996,kenyonErbiumSilicon2005,weissErbiumDopantsNanophotonic2021,gritschNarrowOpticalTransitions2022,berkmanObservingEr32023}, lithium niobate \cite{staudtInterferenceMultimodePhoton2007,sinclairSpectroscopicInvestigationsTi2010,saglamyurekBroadbandWaveguideQuantum2011,duttaIntegratedPhotonicPlatform2020,yangControllingSingleRare2022}, silicon nitride \cite{gongLinewidthNarrowingPurcell2010,gongObservationTransparencyErbiumdoped2010,dingMultidimensionalPurcellEffect2016} or doped thin films grown on silicon \cite{singhEpitaxialErdopedSilicon2020,dibosPurcellEnhancementErbium2022}. Compatibility with standard semiconductor foundry processes is an essential feature for any modality that will be able to scale to the necessary size to enable a quantum internet. For example, it is expected that individual repeater nodes will need on the order of hundreds or thousands of physical qubits for fault-tolerant operation depending on the system metrics and protocols used \cite{durQuantumRepeatersBased1999,muralidharanUltrafastFaultTolerantQuantum2014,muralidharanOptimalArchitecturesLong2016}. This illustrates the size of the challenge to develop a quantum internet that will require extensive networks of repeaters and underscores the need for scalable and repeatable fabrication processes. 

In this review, we discuss the potential for rare-earth ions embedded in a solid-state host to enable a quantum internet through the development of quantum repeaters. We first outline the essential elements of a quantum network, including protocols for the generation, storage, and distribution of entanglement. We then discuss the properties of embedded rare-earth ions in the context of the needs of a quantum network and provide an overview of state-of-the-art developments and outstanding challenges that pertain to the development of repeaters with embedded rare-earth ions. We give particular consideration to erbium, which in its Er$^{3+}$ state is particularly suitable for quantum networking applications and we indicate some near term benchmarks that can be realized by scalable devices for impact in developing quantum networks. Finally, we provide a perspective on how rare-earth ion technologies may further enable the growth of future quantum networks by connecting different qubit modalities using quantum state transduction protocols. These protocols have the power to connect otherwise disparate qubit modalities and will be a boon for the development of global and multi-functional quantum networks.

\section{Quantum networks}\label{sec:QuantumNetworks}
\begin{figure}
    \centering
    \includegraphics[scale=0.75]{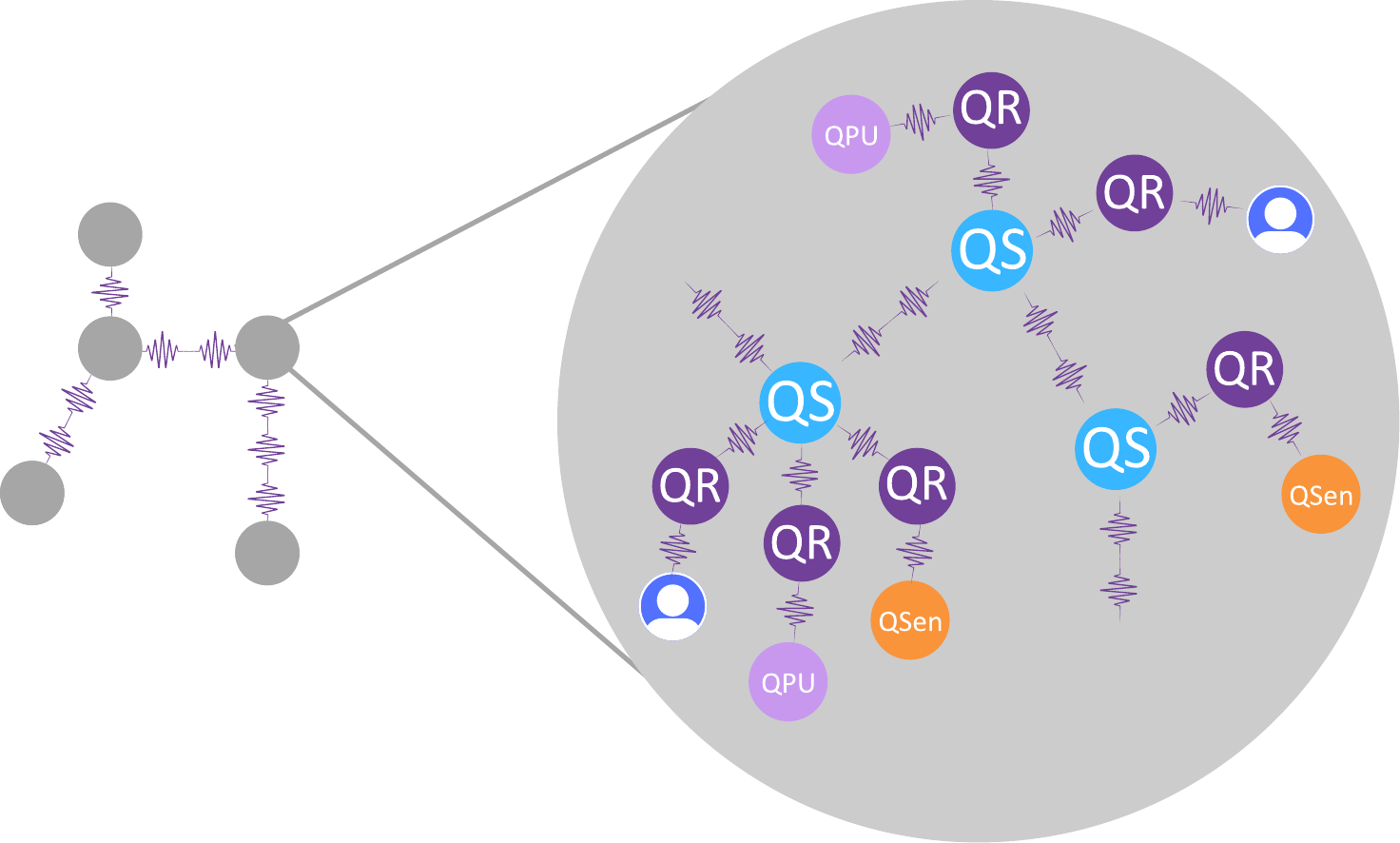}
    \caption{A quantum network where nodes are interconnected via photonic channels. The network comprises multiple length scales where quantum repeaters (QR) and quantum switches (QS) facilitate the distribution of entanglement across local and distant nodes for secure communication between users, quantum processors and computers (QPU), and measurement devices and sensors (QSen).}
    \label{qNetwork}
\end{figure}

The essential function of a quantum network is to distribute information processing efficiently across a web of local quantum computing machines, sensors, and communication devices that are easily controllable, as shown in figure \ref{qNetwork}. The possibility of such a construct however, seems to be determined by a number of conditions that do not appear to be readily compatible with each other if physically realized. Local processing units or nodes are systems where information is processed and stored for long time. Such systems are required to be implemented using quantum hardware that on the one hand exhibit a prolonged coherence time by being well isolated from the surrounding environment, and on the other hand, are highly amenable to controlled interactions with ancillary systems. Moreover, the communication of quantum data across various nodes can only be realized using systems that can traverse long distances with negligible decoherence such as optical photons. Having such weakly-coupling particles interacting with nodes in a controlled way as to mediate information between them seems to be another challenge in building a functional quantum network. Therefore, it seems that the possibility of a quantum network hinges on the ability to realize highly controlled interactions between well isolated nodes such as trapped atomic systems and photons forming a hybrid system of light and matter.

Quantum networks, depending on their task, can be divided into two main categories: computation networks and communication networks. The function of a quantum computation network is to spread entangled states across a large number of nodes that are interconnected by photons to perform large-scale information processing tasks. A modular distributed quantum computer \cite{duanColloquiumQuantumNetworks2010, monroeScalingIonTrap2013} where nodes (logical processing units) are linked through a photonic network is paradigmatic of a quantum computation network, which is capable of supporting universal one-way quantum computation \cite{raussendorfOneWayQuantumComputer2001}. A communication network on the other hand is a scheme to generate a maximally entangled state shared over arbitrary large distances.  In what follows, we will have a quick look at the blueprint of a quantum network within the two aforementioned paradigms.

\textit{Modular quantum computation;} A modular architecture consists of a collection of nodes that are linked through photonic interfaces. Each node is a sequence of qubits that are amenable to fast and deterministic local interactions for realizing single and two-qubit logic gates with error rates sufficiently low to perform fault-tolerant computation. Within each node, the qubits are divided into two classes: memory qubits on which information is processed and stored, and communication qubits that are linked to their counterparts situated within another node through a photonic channel. The function of communication qubits is to facilitate the implementation of two-qubits gates between any pair of memory qubits, one from each node by means of indirect interaction mediated by photons. This architecture allows for the construction of cluster states shared among a large number of nodes offering a platform to run  measurement-based fault-tolerant universal quantum computation \cite{briegelMeasurementbasedQuantumComputation2009}.

{\textit{Quantum communication;}} A quantum communication network is an alternative architecture that can essentially be thought of as a modular quantum computer with a linear topology. The function of a quantum communication network is the distribution of purified maximally entangled states shared exclusively between a pair of nodes separated by an arbitrary large distance. This can be achieved either by constructing a cluster state across the network and performing local gates on intermediary nodes as to distill a maximally bipartite entangled state shared between the end-nodes, or alternatively, by means of entanglement swapping between neighboring nodes in a successive hierarchical fashion \cite{briegelQuantumRepeatersRole1998}.

Central to the functioning of quantum networks discussed above is the capacity to implement what is referred to as \textit{quantum repeater scheme} \cite{briegelQuantumRepeatersRole1998}: a protocol that can \textit{generate, store} and \textit{distribute} entanglement amongst pairs of spatially separated quantum nodes \textit{efficiently}. That is, the key to a realistic quantum network is the ability to perform the following three steps with arbitrary degree of accuracy, and with a polynomial overhead in the resources: (i) generating entanglement between the communication qubits mediated by photons, (ii) transferring the induced entangled state onto the memory qubits for storage, and (iii) connecting and purifying the stored entangled states across the network through local operations. In what follows, we provide a minimal review on the schemes that realize these three steps.

\subsection{Entanglement generation}\label{sec:ent_generation}

\begin{figure}[!h]
    \centering
    \includegraphics[scale=0.15]{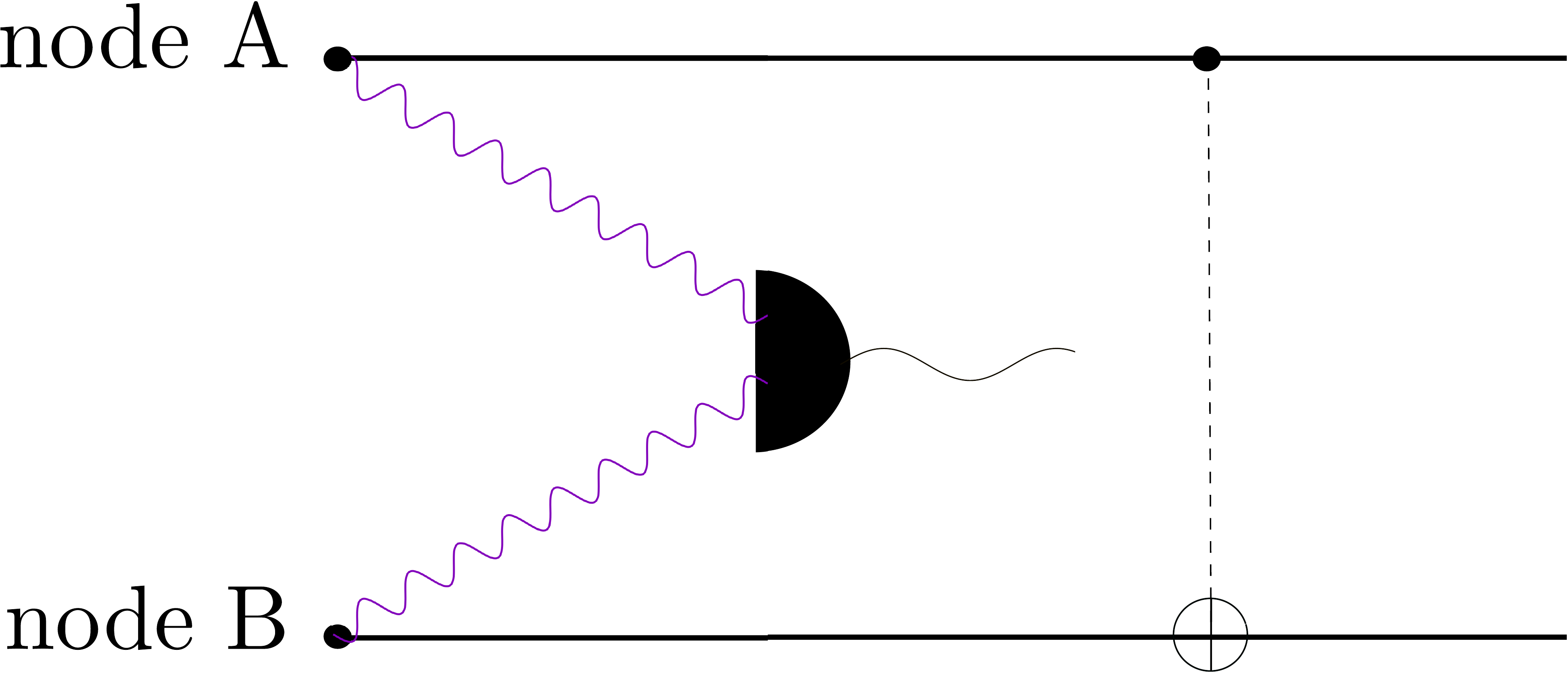}
    \caption{ At each node a qubit-photon pair is generated. Photons are collected for a Bell-state measurement. Detection of a single photon would herald the generation of an entangled state shared between the nodes.}
    \label{ES}
\end{figure}

The core algorithm underlying the generation of entanglement between a pair of nodes is illustrated in figure \ref{ES}: first an entangled qubit-photon pair is generated within each node, and then a Bell-state measurement of the collected photons projects the nodes onto an entangled state. We will briefly summarize the two main protocols that realize this entangling circuit.

\subsubsection{Single emitters}

\begin{figure}
    \centering
    \includegraphics[scale=0.2]{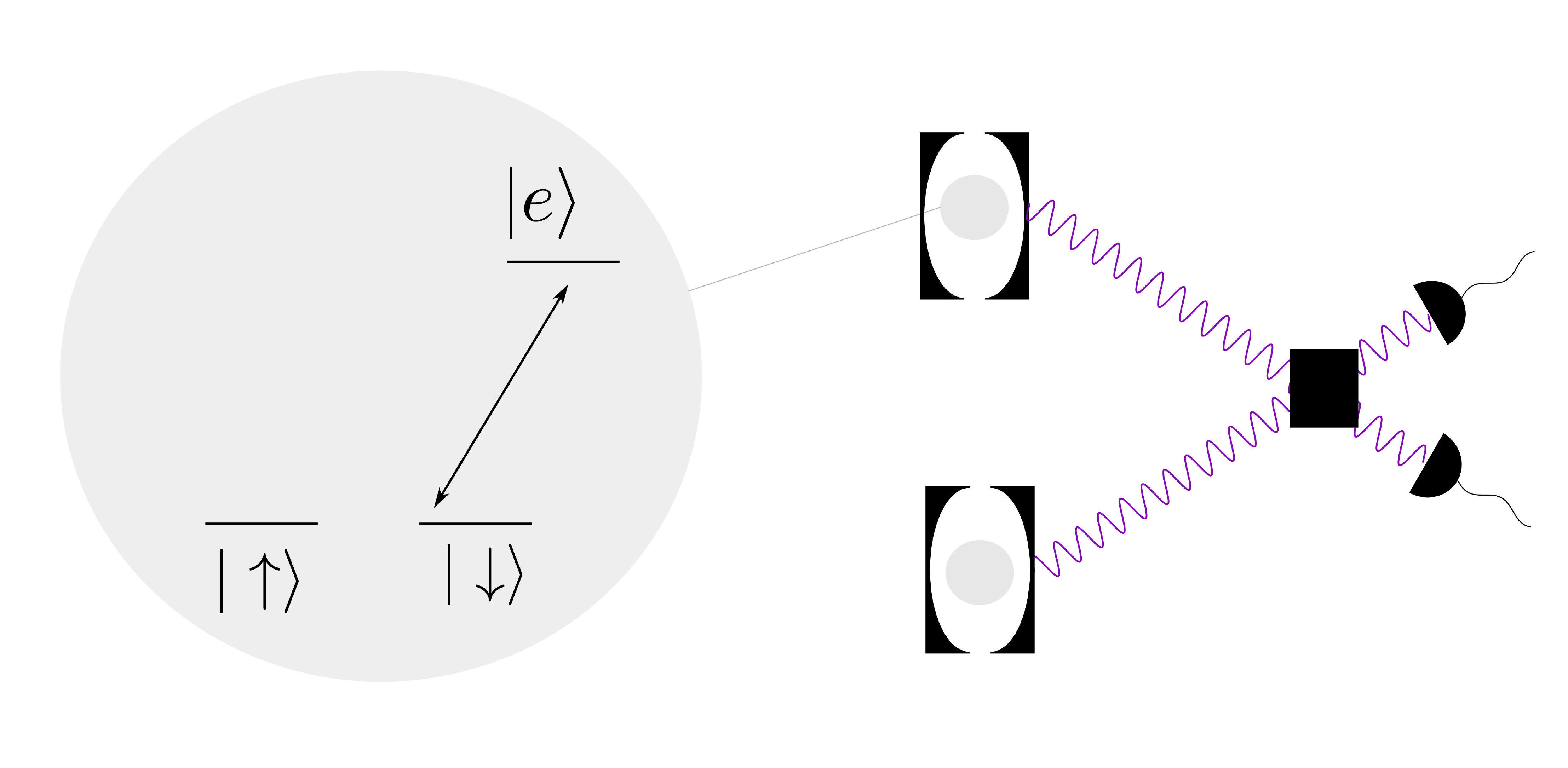}
    \caption{In each cavity a three-level system (a qubit system $\{\vert \uparrow\rangle,\vert\downarrow\rangle\}$ plus an excited state $\vert e\rangle$) is situated. Driving the $\vert\downarrow\rangle\rightarrow\vert e\rangle$ transition, followed by the emission of a photon brings the excited state back to $\vert\downarrow\rangle$. Conditioned on the observation of a single photon at the detection outputs, the entire state is projected onto an entangled qubit state. }
    \label{EScheme1}
\end{figure}

In this approach \cite{boseProposalTeleportationAtomic1999, barrettEfficientHighfidelityQuantum2005}, each node consists of an atomic system with an effective three-level Lambda configuration - which comprises two relatively stable low lying states $\{\vert\uparrow\rangle, \vert\downarrow\rangle \}$ encoding the qubit state, and an excited state $\vert e\rangle$ - situated within an optical cavity, which is schematically depicted in figure \ref{EScheme1}. 
Both of the qubit systems are initially prepared in a superposition state forming the product state $[\frac{1}{\sqrt{2}}(\vert \uparrow\rangle + \vert \downarrow\rangle)]^{\otimes 2}$. Next, each qubit is driven by a control pulse that is resonant with the $\vert\downarrow\rangle \rightarrow\vert e\rangle$ transition. From this moment on, the effective dynamics of the system is governed by the Jayne-Cummings Hamiltonian:
\begin{eqnarray}
H = \sum_{i=1}^2 g_i(\sigma_i^- a_i^{\dagger} + \sigma^+_i a_i),
\label{BK}
\end{eqnarray}
where $g_i$ denotes the coupling between the qubit and the cavity mode, and the $\sigma_i^{\pm}$ ($\{a, a_i^{\dagger}\}$) correspond to the creation and annihilation operators associated with the qubit (cavity photon). As the Hamiltonian \ref{BK} indicates, the transition $\vert\downarrow\rangle \leftrightarrow\vert e\rangle$ is coupled to the cavity mode, which after a short period of time causes the qubit to undergo a spontaneous emission that results in a locally entangled state shared by the qubit and the emitted photon in the cavity resulting in the state $\vert \psi\rangle =[ \frac{1}{\sqrt{2}}(\vert \downarrow 1\rangle +\vert \uparrow 0\rangle)]^{\otimes 2}$. The emitted photons are then collected from the cavities and are guided towards a beam splitter just before being measured in an optical detection output. The function of the beam splitter is to erase the information regarding the trajectory that each photon has taken, so that the detection of a single photon could have originated from either of the nodes. In this way, conditioned on the observation of a single photon, the collective state of the qubit-photon pair is updated under the action of the emission operator $l_{\pm}=\frac{1}{\sqrt{2}}(a_1 \pm a_2)$:
\begin{eqnarray}
l_{\pm}\vert\psi\rangle \rightarrow \vert\phi_{\pm}\rangle + \vert\zeta_{\pm}\rangle,
\label{BK2}
\end{eqnarray}
where $\vert\phi_{\pm}\rangle \propto (\vert\uparrow\downarrow\rangle\pm\vert\downarrow\uparrow\rangle)\vert 00\rangle $ denotes the desired maximally entangled state between the qubit pair, and $\vert \zeta_{\pm}\rangle \propto (\vert 01\rangle\pm\vert 10\rangle)\vert \downarrow\downarrow\rangle$. As the equation \ref{BK2} indicates, there exists the possibility of a second photon being observed, which is encoded in $\vert \zeta_{\pm}\rangle$. However, due  to experimental imperfections such as photon loss and limited efficiency of the detectors, the state $\vert \zeta_{\pm}\rangle$ is, as well, compatible with the single photon detection with a non-vanishing probability. That is, the residual entanglement between the qubit-pair and the photon-pair can not be removed due to imperfect knowledge of the system, which leads to the decoherence of the qubit-pair into a mixed state
\begin{eqnarray}
\rho = p_1 \vert u_1^{\pm}\rangle\langle u^{\pm}_1\vert + p_2 \vert u_2\rangle\langle u_2\vert
\label{BK3}
\end{eqnarray}
where $\vert u_1^{\pm}\rangle =\frac{1}{\sqrt{2}}  (\vert\uparrow\downarrow\rangle\pm\vert\downarrow\uparrow\rangle)$, and $\vert u_2\rangle = \vert \downarrow\downarrow\rangle$. The way to suppress the second term in the equation \ref{BK3} is to flip the qubit states, $\vert\downarrow\rangle \leftrightarrow \vert \uparrow\rangle$ and repeat the previous procedure. In this way, the state $\vert\downarrow\downarrow\rangle$ is turned to $\vert\uparrow\uparrow\rangle$, which cannot contribute to any emission, whereas the state $\vert u_1^{\pm}\rangle$ would still emit a single photon. Therefore, the observation of a single photon per each cycle would herald the generation of a pure entangled state $\vert u_1^{\pm}\rangle$ among the qubit-pair.\\
\\
The fidelity of this entangling protocol can be affected by various sources of error, of which two main categories are the decoherence of the qubit systems and the reduced indistinguishability of the photon pairs \cite{reisererCavitybasedQuantumNetworks2015}. Decoherence of the qubit due to its unavoidable coupling to the environment leads to the reduction in the initial entanglement measure between the qubit and the photon, which has a direct detrimental impact on the amount of the ultimate entanglement induced between the qubit-pair. The distinguishability between the photons is another source of infidelity as the functioning of the protocol hinges on the inability of the measuring apparatus to discern from which node the photon arrives. This condition requires accurate control over the qubit-photon interaction, photon emission, frequency and polarization.

\subsubsection{Atomic ensembles}

One of the key requirements for single emitters to function as communication qubits is having them situated within optical resonators of very high quality. An alternative approach that bypasses this challenge is to use an ensemble of single emitters, which greatly enhances the control over photon generation due to their collective effect. A remarkable entangling scheme that is based on the atomic ensemble, and which we refer to as the DLCZ protocol \cite{duanLongdistanceQuantumCommunication2001} involves a pair of ensembles each described effectively as a collection of three-level Lambda systems with two grounds state $\{\vert\uparrow\rangle, \vert\downarrow\rangle \}$, and an excited state $\vert e\rangle$. Each ensemble is prepared at the ground state before being driven by a radiation pulse that is off-resonant on the transition $\vert\downarrow\rangle \leftrightarrow\vert e\rangle$, which leads to a spontaneous Raman emission $\vert e\rangle \leftrightarrow\vert\uparrow \rangle$. This leads to the generation of a Stokes photon and a single atomic excitation (logical state $\vert\boldsymbol{\Uparrow}\rangle$) smeared over the entire ensemble:
\begin{eqnarray}
\vert\boldsymbol{\Uparrow}\rangle = \frac{1}{\sqrt{N}} \sum _{i=1}^N \alpha_i \vert \downarrow\rangle_1\vert \downarrow\rangle_2...\vert \uparrow\rangle_i...\vert \downarrow\rangle_N
\label{ensembleState}
\end{eqnarray} 
where the phase factor $\alpha_i$ depends on the radiation mode and the position of the $i$th atom, and $N$ denotes the total number of the atoms within the ensemble. The interaction between the emitted photon and the ensemble described above is effectively captured by the Hamiltonian:
\begin{eqnarray}
H = g(s^{\dagger}a^{\dagger}+sa)
\label{DLCZHamiltonian}
\end{eqnarray}
where $g$ denotes the coupling constant that is a function of the pulse intensity, number of atoms within the ensemble and the strength of the Raman transition, and $a$ and $s$ denote the annihilation operator for Stokes photon and the atomic excitation respectively. The evolution induced by this Hamiltonian generates an ensemble-photon pair entangled state that can be written as a series expansion for small values of $gt$:
\begin{eqnarray}
e^{-iHt}\vert\boldsymbol{\Downarrow}\rangle\vert 0\rangle \rightarrow \vert\boldsymbol{\Downarrow}\rangle\vert 0\rangle + ip^{\frac{1}{2}}\vert\boldsymbol{\Uparrow}\rangle\vert 1\rangle + O(p)
\label{DLCZEvolution}
\end{eqnarray}
where $p = (gt)^2$ denotes the probability for a single Stokes photon emission. At this stage, in a similar fashion to the single-emitter scheme discussed above, the Stokes photons are collected from both of the ensembles, and sent into a beam-splitter as to erase the which-way information of the photons before being observed at the detection outputs. Conditioned on a single photon being detected, the state of the entire system is updated under the spontaneous emission operator $l_{\pm}=\frac{1}{\sqrt{2}}(a_1 \pm a_2)$ that induces the desired entangled state shared between the ensemble pair:
\begin{eqnarray}
l_{\pm}\big(\vert\boldsymbol{\Downarrow}\rangle\vert 0\rangle + ip^{\frac{1}{2}}\vert\boldsymbol{\Uparrow}\rangle\vert 1\rangle\big)^{\otimes 2}\rightarrow p^{\frac{1	}{2}}\big(\vert\boldsymbol{\Downarrow}\boldsymbol{\Uparrow}\rangle + i\vert\boldsymbol{\Uparrow}\boldsymbol{\Downarrow}\rangle\big)\vert00\rangle + O(p).
\label{DLCZProtocol}
\end{eqnarray}
As equation \ref{DLCZProtocol} indicates, the generated entangled state is not pure due to the presence of multi-photon emission with the probability $	O(p)$. In order to suppress these higher order terms one needs to work with very low emission rate $p$, which in return, reduces the probability for generating the entangled state itself, and subsequently, prolonging the protocol duration . Therefore, to be able to distribute the prepared entangled state across a network via the swapping operation, it is necessary for the ensemble-pair system to have a long coherence time for being able to store the heralded entanglement while the adjacent pair is undergoing the protocol.

\subsection{Entanglement storage}\label{sec:storage}
Once the generation of entanglement between a pair of nodes is heralded, the next step is to map the shared state onto a quantum memory for storage and processing. Depending on the function that a quantum network performs the state mapping process can be divided into two separate classes: (1) for a quantum computation network where each node consists of a sequence of single emitters, the state is transferred from a communication qubit onto a memory qubit with long coherence through controlled local unitary gates at each node; (2) for a quantum communication network however, the atomic ensembles function both as communication and memory qubit and therefore the state-mapping procedure is no longer needed. 

As we discussed earlier single emitters are a prime candidate for realizing distributed quantum information processing due to their high degree of controllability both in terms of their coupling to photonic channels, and their interaction with neighboring counterparts for unitary logic operations. One of the challenges linked with single-emitter-based architecture however, is to eliminate cross-talk between the interactions that a communication qubit undergoes with the photonic interface and with the memory qubits \cite{monroeScalingIonTrap2013}. This means that a communication qubit must be well isolated from the rest of the node in order to prevent the unwanted excitation of the memory qubits, since such excitation caused by the scattered photons can introduce irreversible error channels in the computational tasks running within the node. There are two ways to effectively shield the memory qubits from the scattered photons. One way is to spatially separate the communication qubit from the rest of the node during the photon exchange. In this scheme, the communication qubit is physically transported to an optical cavity situated at a safe distance from the memory qubits for entanglement operation. Once the entanglement is established, the communication qubit is brought back to the vicinity of the other qubits for the state transfer operation via local unitary gates. Despite the effective isolation that such scheme offers, it requires a highly precise manipulation of the communication qubit during the transport, which leads to an increase in the control overhead. An alternative way to suppress such cross-talks is to utilize different atomic species with distinct transition frequencies for communication and memory qubits. In this way the photon emitted from a communication qubit would simply be invisible to the rest of the node as its frequency is sufficiently off with respect to the memory qubits' resonance frequency. Moreover, the division of each node into different atomic species allows for more flexibility in the choice of qubits since unlike the memory qubits, it is not necessary for the communication qubtis to have long coherence characteristic, for once the entanglement between them is induced their state can be promptly written onto the memory qubits.

\begin{figure}[!h]
    \centering
    \includegraphics[scale=0.4]{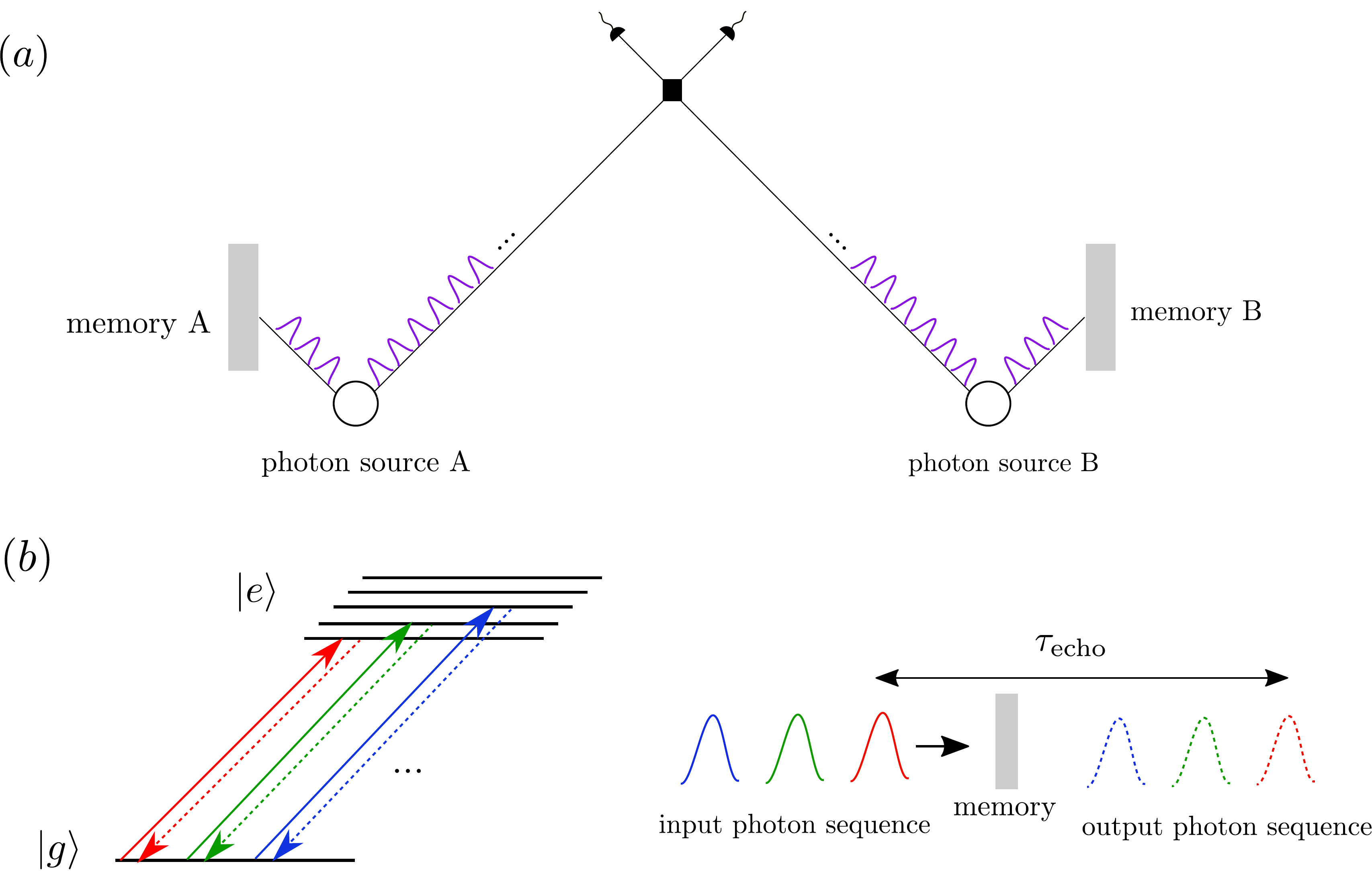}
    \caption{ $a)$ Entanglement generation scheme using temporal multi-mode memories: The photon sources at sites $A$ and $B$ where the memories are located can each emit an entangled photon pair into a sequence of distinct temporal modes where for each mode one is sent to the detection output and the other one is stored in the memory. The detection of a single photon behind the beam splitter would project the memories into an entangled state.  The source can be triggered a large number of
times in every communication time interval. $b)$ Temporal multi-mode memory system based on photon echo principle: a sequence of photons are absorbed and re-emitted by a comb-structure profile that is induced though inhomogeneous broadening of the spectra linked with the atomic ensemble memory. }
    \label{AFC}
\end{figure}

The alternative design for quantum nodes in a communication network involves atomic ensembles \cite{wangFieldDeployableQuantumMemory2022,caoEfficientReversibleEntanglement2020} which as we discussed above exhibit deterministic interactions with photons without utilizing any cavity enhancement techniques thanks to their the collective effect. A salient aspect of atomic-ensemble-based quantum memories is their multi-modal storage capacity \cite{sangouardQuantumRepeatersBased2011}. In the entanglement creation protocols discussed above, the heralding of the established entanglement entails the emitted photon traveling to the measurement output and the result of a positive detection traveling back to the node. In this way, the overall rate at which such protocol can be repeated is bounded by the communication time $\tau \propto L/c $ where $L$ denotes the photonic channel length. To overcome this limitation schemes that utilize multi-mode memory systems that can store large numbers of distinguishable photonic modes have been proposed \cite{sangouardAnalysisQuantumMemory2007, afzeliusMultimodeQuantumMemory2009, sangouardQuantumRepeatersBased2011}. More concretely, using quantum memories that can store $N$ distinct modes would allow for the $N$-times repetition of the entangling protocol per communication time, which improves the entanglement generation rate by a factor of $N$, and therefore the requirement for a quantum memory to have extremely long coherence time can be relaxed by the same factor. Various types of degrees of freedom can be chosen as to form the basis for these modes such as spatial or spectral modes of the photons. A desirable option however, is to utilize the distinguishable temporal modes or the time-bin degrees of freedom linked with the photon, which is also naturally suited for the transmission of information used in current telecommunication networks. The main entangling scheme that is well adapted for temporal multiplexing is a variation of the DLCZ protocol where the entanglement generation and storage are separated. In this approach \cite{sangouardAnalysisQuantumMemory2007}, photon pair sources and absorptive memories are integrated as to emulate a sequence of DLCZ operations per communication time as illustrated in figure \ref{AFC}.

The entangling scheme summarized above is based on memory systems that offer storage in various distinguishable temporal modes $\{\vert 0\rangle_i, \vert 1\rangle_i\}$. There are varied methods to realize these temporal multi-mode memories, a class of which - referred to as photon echo principle - is well suited for atomic ensembles in solids such as crystals doped with rare-earth ions \cite{lago-riveraTelecomheraldedEntanglementMultimode2021}.  The essence of the techniques that are based on photon echo principle is to preserve the one-to-one temporal correspondence between the absorption and the emission of each photon so that photons that are absorbed at separate moments are emitted at separate moments. The storage schemes that are based on echo techniques utilize reversible absorption of a single mode by a media the spectrum of which is subject to inhomogeneous broadening. Once the single photon is absorbed the memory state is mapped onto a collective atomic excitation :
\begin{eqnarray}\label{eqn:AFC_collective}
\vert 1\rangle = \sum_k c_ke^{i\delta_k t}\vert g_1\rangle\vert g_2\rangle...\vert e_k\rangle...\vert g_N\rangle
\end{eqnarray}
where $\delta_k$ denotes the detuning of the $kth$ atom from the photon's central frequency. The core idea behind the echo technique is to steer the detuning spectrum such that after a certain period of time all the coefficients $e^{i\delta_k t}$ realign causing the memory state to re-phase. As a consequence the absorbed photon will be emitted in a well defined temporal mode thanks to the collective interference effect amongst all the emitters within the atomic ensemble. In this way the information regarding the temporal relation between a sequence of photons is encoded in the relative phases of the atomic excitations at different frequencies. As a result, a sequence of photons that are absorbed by the memory at different moments will be emitted at distinct instants of time each separated (from the absorption time) by a constant duration determined by the controlled re-phasing period.

\subsection{Entanglement distribution}\label{sec:distribution}
Once the entangled state is stored among pairs of nodes, the next step is to link these pairs as to share the entanglement between the nodes situated at large distances from each other that are not directly connected via photon channels. This can be done by performing an entanglement swapping operation \cite{sangouardQuantumRepeatersBased2011} on the nearby nodes in a heralded fashion mapping the distant node-pair into an entangled state. The swapping operation consists of a Bell-state measurement on the intermediary nodes, and conditioned on the result the composite state of the end-point nodes is projected onto an entangled state. Let us imagine that two pairs of nodes $A-B$ and $C-D$ are each prepared in an entangled state given in (\ref{DLCZProtocol}). The information stored within the memories $B$ and $C$ is converted into photons propagating in well defined direction, and subsequently collected at a beam splitter as to erase their which-way information before being measured at detection outputs. Conditioned on the observation of a single photon the global state is updated under the emission operator $\frac{1}{\sqrt{2}}(a_B+a_C)$ projecting the memories $A$ and $D$ into an entangled state. The dominant source of error in a swapping operation is related to photon loss due to inefficiencies linked  with detectors and memories. That is the measuring outputs can erroneously signal a single photon detection while having two photons stored in memories $B$ and $C$, which leads to the presence of an additional vacuum component in the generated entangled state between $A$ and $D$:
\begin{eqnarray}
\rho_{AD} = p_1\vert u_1^{\pm}\rangle\langle u_1^{\pm}\vert + p_2\vert u_2\rangle\langle u_2\vert
\end{eqnarray} 
where $\vert u_1^{\pm}\rangle = \frac{1}{\sqrt{2}}(\vert \uparrow \downarrow\rangle_{AD} \pm\vert \downarrow\uparrow\rangle_{AD} )$, and $\vert u_2\rangle = \vert \downarrow\downarrow\rangle_{AD}$. In order to connect a pair of distant nodes separated by $N$ links ($N-1$ intermediary nodes) one needs to perform the swapping operation $N$ times, and with each such connection the ratio $p_1/p_2$ decays. This leads to a rapid decrease in the fidelity of the induced entangled state shared between the end-point nodes such that at large $N$ it will not be possible to error-correct the state using any purification protocol. To bypass this difficulty an error correction scheme called the nested purification protocol \cite{briegelQuantumRepeatersRole1998} has been proposed, which consists of connecting and purifying in a successive hierarchical fashion. The fundamental principle underlying this protocol is the entanglement distillation that makes it possible to purify an entangled state to arbitrary degree of accuracy from an ensemble of noisy imperfect entangled pairs using only local operations at each node. The nested purification protocol can be carried out by iterating the following two steps: first applying the swapping operation to a small number of links $L<<N$, producing  an ensemble of noisy entangled pairs, and second performing local single and two-qubit gates on the connected nodes as to distill a smaller number of highly entangled states with suppressed error rates. It can be easily checked that if the number of elementary imperfect entangled states necessary to purify an state across $L$ links is $M$, then the total number of such elementary states required for a network of $N = L^n$ links is $R= (LM)^n$. In other words, the overhead in the resources scales polynomially with $N$ as $R$ can be re-written as $R = N^{\log_L M+1}$. $M$ can be determined by a number of factors including the fidelity of the elementary imperfect raw entangled pairs, the network platform, the type of the distillation protocol used, and the nature of the  dissipative processes involved in the operations of swapping and purification.

\section{Rare-earth ion doped materials for quantum networking}
The focus of this section is to describe the properties and capabilities of rare-earth ion doped materials in terms that are useful for constructing quantum communication networks as discussed in section \ref{sec:QuantumNetworks}. Rare-earth ions can interface with a range of optical wavelengths from the visible to the telecom, and also possess controllable and long lived spin states that make them compatible with many protocols and architectures for building out a quantum network. In this section, we aim to characterize these properties in terms of how they relate to the construction of quantum networks. We will focus on the ways rare-earth ion doped materials have been used for quantum resource generation, quantum resource storage, and quantum resource distribution. We will also outline some of the challenges that must be overcome for rare-earth ion doped materials to develop into a platform that can support large scale quantum networks.

\begin{figure}
\centering
\includegraphics[scale = 0.3]{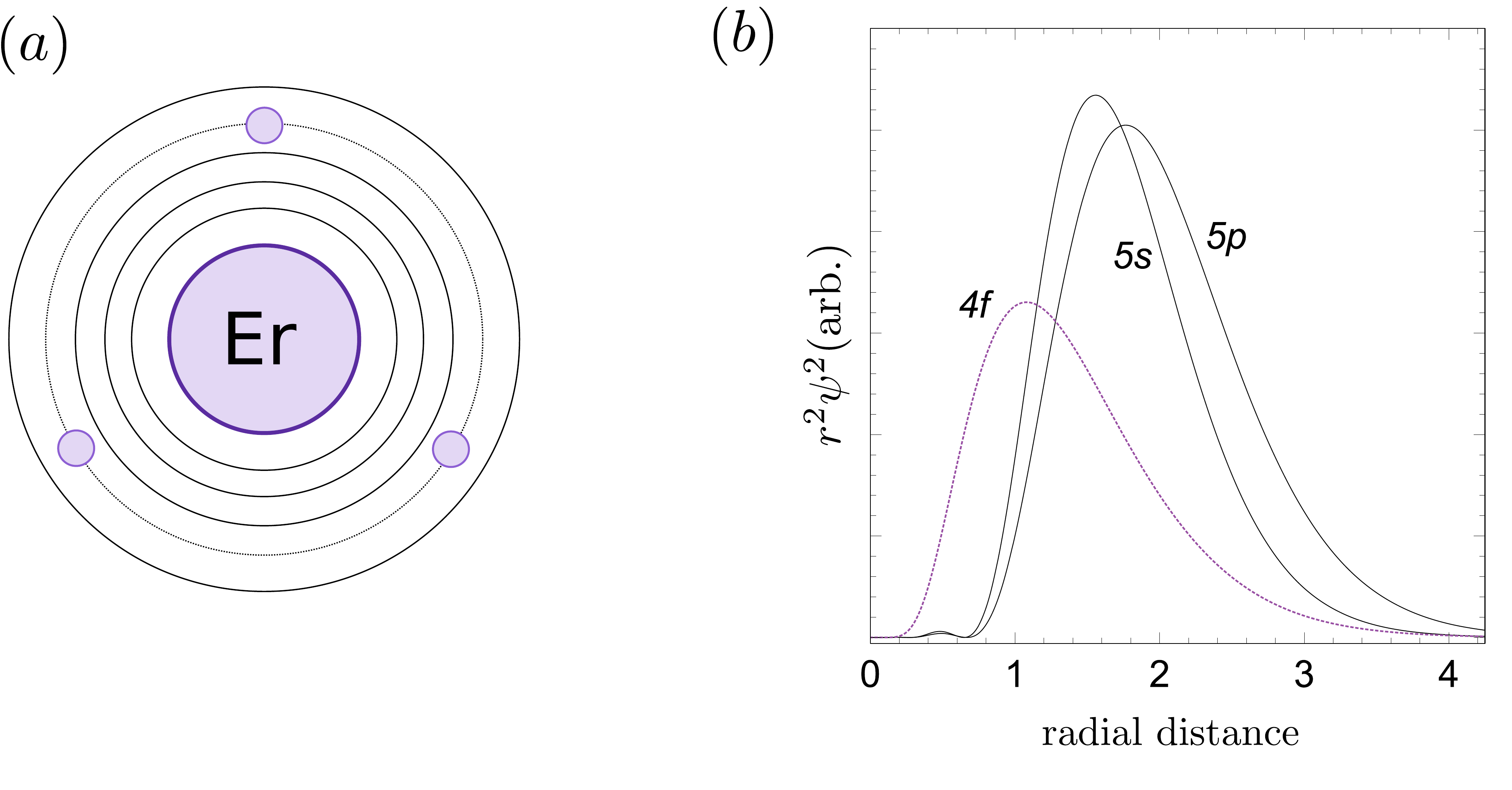}
    \caption{(a) Simplified picture of a rare-earth ion showing only the unpaired electrons orbiting the ion nucleus. Orbitals are drawn for the first five principle quantum numbers. Triply ionized erbium (Er$^{3+}$) is shown, having three unpaired electrons in the $4f$ orbital. (b) Radial wavefunctions for the outer orbitals of Er$^{3+}$ as a function of distance from the nucleus. The unfilled $4f$ orbital is shielded by the filled $5s$ and $5p$ orbitals.}
    \label{fig:simple_valence}
\end{figure}

Rare-earth elements are a grouping of metals in the periodic table that in their readily formed triply ionized state possess a partially filled $4f$ orbital shielded by filled $5s$ and $5p$ outer orbitals. The configuration of the unpaired electrons in the $4f$ orbital of Er$^{3+}$ is illustrated in figure \ref{fig:simple_valence}. The shielding of the $4f$ orbital allows the energy levels of the ion to experience only a slight and well understood perturbation when doped into a crystalline environment \cite{diekeSpectraDoublyTriply1963,macfarlaneHighresolutionLaserSpectroscopy2002,thielRareearthdopedMaterialsApplications2011}. The unpaired electrons within the $4f$ orbital provide a variety of low lying spin levels that can be used as qubit states as well as excited electronic levels that can couple to optical photons. These properties underlie all the applications in quantum networking which are outlined in this section.  An established understanding of the properties of both the ions and the chosen host crystals makes it possible to turn our attention to the goal of engineering devices that can then enable the development of quantum networks. 

\subsection{Quantum resource generation}\label{sec:resource_generation}
\subsubsection{Considerations for single photon sources}\label{sec:considerations_spe}
In the context of using rare-earth ions for quantum networking, we will consider quantum resource generation to refer to the production of single photons that can be used within a network. In accordance with the protocols outlined in section \ref{sec:ent_generation}, these photons can then be used to generate or distribute entanglement throughout the network. The primary considerations for judging the quality of a single photon source as a resource for quantum networking are the brightness, coherence time, and indistinguishability between the emitted single photons. Indistinguishability between photons is a measure that determines the degree to which the emitted photons are identical. It encompasses every property of a photon including wavelength and polarization, as well as spatial and temporal mode profile. The degree of indistinguishability between photons is the most important feature to consider for a single photon source to be used in a quantum network since photons must be indistinguishable to distribute entanglement, as highlighted in section \ref{sec:QuantumNetworks}. The source brightness determines how many single photons can be inserted into the network over a given time frame, and is closely related to the single photon emission probability which quantifies the errors that may occur due to undesired multi-photon emission events. In this regard, atomic and atom-like sources of single photons offer the advantage of near-deterministic operation \cite{kuhnDeterministicSinglePhotonSource2002,aharonovichSolidstateSinglephotonEmitters2016}, while sources based on spontaneous parametric down conversion \cite{guoParametricDownconversionPhotonpair2017} or spontaneous four-wave mixing \cite{paesaniNearidealSpontaneousPhoton2020} rely on heralding to suppress multi-photon events. In general, sources with both a high brightness and a high single photon emission probability are required for different applications. Finally, the coherence time determines how long a photon can travel in the network and still interfere with another identical photon. In this section we outline how each of these properties are related to one another before discussing the progress that has been made so far in building single photon sources from rare-earth ions in solid-state media.

In the solid state, the properties of a single photon emitter are intimately related to the host material it is embedded in. A single photon source has an emission rate, $\gamma$, that is inversely related to the optical excited state lifetime of the emitter through the relationship $\gamma = 1/T_1$, where $T_1$ is the lifetime. Shorter excited state lifetimes therefore provide larger emission rates in the absence of non-radiative decay channels. If non-radiative decay channels exist, the total decay rate is the sum of the radiative and non-radiative parts, $\gamma = \gamma_r + \gamma_{nr}$, where $\gamma_r$ and $\gamma_{nr}$ are the radiative and non-radiative rates, respectively. An optically excited ion may then have a pathway to relax back to the ground state, or another state altogether, without emitting a photon in the desired mode. While these additional decay pathways may find use in control or read-out protocols \cite{wolfowiczQuantumGuidelinesSolidstate2021}, they are generally disadvantageous in applications that require the emission of single photons into a single optical mode. Therefore our goal is to limit the presence of non-radiative decay channels by choosing emitter and material combinations that minimize them \cite{bassettQuantumDefectsDesign2019}. We will also see in section \ref{sec:rei_enhancement} that there are other tools, in addition to emitter and host crystal choice, that allow us to improve the radiative properties single photon emitters. The brightness of the single photon source therefore depends critically on the radiative decay rate of the emitter, as well as how efficiently the emitted photons can then be coupled to the network. In order to achieve a high single photon emission probability, multi-photon emission can be suppressed by tailoring the excitation pulse duration to be sufficiently short with respect to $T_1$ to avoid re-excitation of the emitter \cite{hanschkeQuantumDotSinglephoton2018}.

\begin{figure}
    \centering
    \includegraphics[scale=0.3]{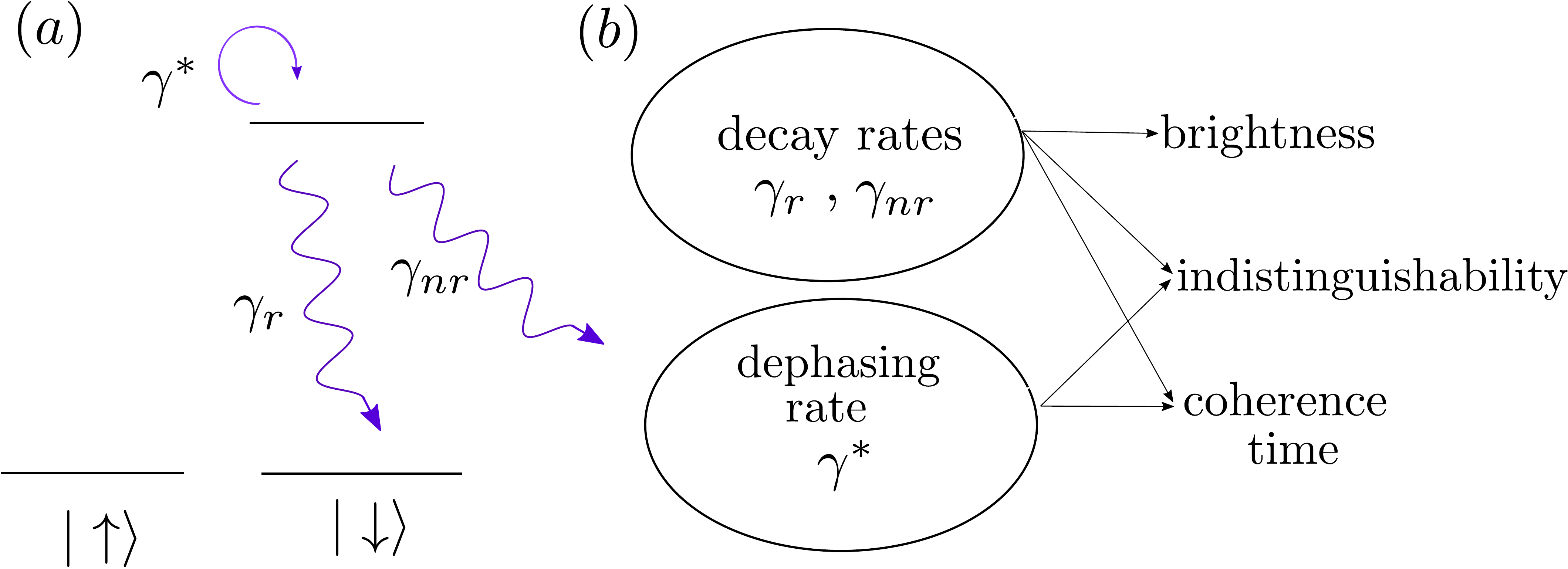}
    \caption{(a) A single photon emitting source subject to decay and dephasing mechanisms. (b) Outline of how decay and dephasing mechanisms are related to the brightness, indistinguishability, and coherence time of a single photon source.}
    \label{fig:decay_rates}
\end{figure}

When a photon is emitted, the coherence time of the photon is given by the Fourier transform limited value $T_2=2T_1$ if no additional dephasing mechanisms are present. Therefore an unavoidable trade-off between emission rate and coherence time exists. This trade-off must be considered when choosing a single photon source for a network, as the coherence time of the source can also effect the achievable spacing of the network nodes. Additionally, the host material may introduce optical dephasing mechanisms from interactions with phonons or fluctuating electric and magnetic fields \cite{bottgerOpticalDecoherenceSpectral2006}. These mechanisms can be incorporated into a dephasing rate, $\gamma^*/2=1/T_2^*$, where each dephasing mechanism has its own characteristic timescale, $T_2^*$. In the presence of this additional dephasing, the coherence time of the emitted photons becomes

\begin{equation*}
    \frac{1}{T_2} = \frac{1}{2T_1} + \frac{1}{T_2^*},
\end{equation*}

\noindent
so the presence of dephasing reduces the coherence time of the emitted photons. The decay and dephasing rates for a single photon emitter are illustrated in figure \ref{fig:decay_rates}a.

Dephasing also reduces the indistinguishability between single photons. The indistinguishability is given by the ratio $I=\gamma/(\gamma + \gamma^*)$, so it is desirable to reduce optical dephasing as much as possible. The indistinguishability between photons emitted by a single photon source can be measured by performing a two photon interference experiment where successive photons from the single photon source are interfered on a beamsplitter. If both photons are identical then they will coalesce and exit the final beamsplitter in the same path, an effect known as Hong-Ou-Mandel interference \cite{loudonMultimodeContinuousmodeQuantum2000}. If the photons are distinguishable, they will be able to exit the beamsplitter along separate paths. In a coincidence counting measurement, the probability of measuring simultaneous clicks between the two detectors is given by $P_{sim}\propto 1-I$, providing a way to directly characterize the indistinguishability between single photons. We summarize the interplay of the decay and dephasing rates on the brightness, coherence time, and indistinguishability between the emitted single photons of the source in figure \ref{fig:decay_rates}b.

\subsubsection{Rare-earth ion single photon emission}\label{sec:rei_emission}
Rare-earth ions embedded into solid-state hosts possess intra-$4f$ optical transitions that become weakly dipole allowed due to the perturbation of the crystalline lattice. These transitions span a wide range of optical wavelengths depending on the chosen ion and experience minimal electron-phonon coupling due to the relative proximity and confinement of the electrons around the ion nucleus \cite{thielRareearthdopedMaterialsApplications2011}. In fact, the intra-$4f$ optical transitions of rare-earth ions are some of the most pristine known optical transitions in the solid-state. Trivalent erbium (Er$^{3+}$) is of particular note for quantum networking because it provides optical transitions in the desirable telecom C-band (1530-1565 nm) that experiences the lowest loss for propagation through optical fiber. For example, the lowest energy optical transition for Er$^{3+}$ ($^4I_{13/2}\rightarrow ^4I_{15/2}$) has an optical wavelength of approximately 1550 nm. Operating directly in the telecom window eliminates the need for optical frequency conversion \cite{kumarQuantumFrequencyConversion1990,huangObservationQuantumFrequency1992,radnaevQuantumMemoryTelecomwavelength2010,guoOnChipStrongCoupling2016,bockHighfidelityEntanglementTrapped2018,singhQuantumFrequencyConversion2019}, which is challenging to achieve in the quantum regime and places extra demands on the overhead required to operate quantum networks.

The weakly allowed $4f$ - $4f$ optical transitions of rare-earth ions have long excited state lifetimes, $T_1$, on the order of 10 ms that lead to long optical coherence times. While these long coherence times are desirable for enabling the distribution of entanglement over long distances, the long lifetimes also consequently lead to a low brightness for individual ions as outlined in section \ref{sec:considerations_spe}. These low photon emission rates present an immediate challenge for utilizing rare-earth ions as a platform for quantum resource generation by making the optical isolation of single ions particularly difficult. 

Initial experiments suggested that isolation of single ions in a crystalline medium could be feasible \cite{langeObservationSingleImpurity1988a,rodrigues-herzogOpticalMicroscopySingle2000,bartkoObservationDipolarEmission2002,malyukinSingleionFluorescenceSpectroscopy2003}, but conclusive optical isolation of single ions was not established until more recently \cite{kolesovOpticalDetectionSingle2012,yinOpticalAddressingIndividual2013,nakamuraSpectroscopySinglePr2014,siyushevCoherentPropertiesSingle2014,utikalSpectroscopicDetectionState2014}. However, these demonstrations of single-ion optical isolation did not collect photons from the lowest lying intra-$4f$ transitions. Rather, they utilized either non-optical methods of detection or optical transitions to higher energy electronic excited states that possess shorter lifetimes. Shorter excited state lifetimes make the higher energy transitions inherently brighter and therefore easier to detect, but also preclude the possibility of exploiting the desirable properties of the lowest lying excited states. Generating single or entangled photons from these transitions is a critical task for building out long distance quantum networks with rare-earth ion doped materials.

\subsubsection{Cavity-enhanced optical emission}\label{sec:rei_enhancement}
\begin{figure}
    \centering
    \includegraphics[scale=1.0]{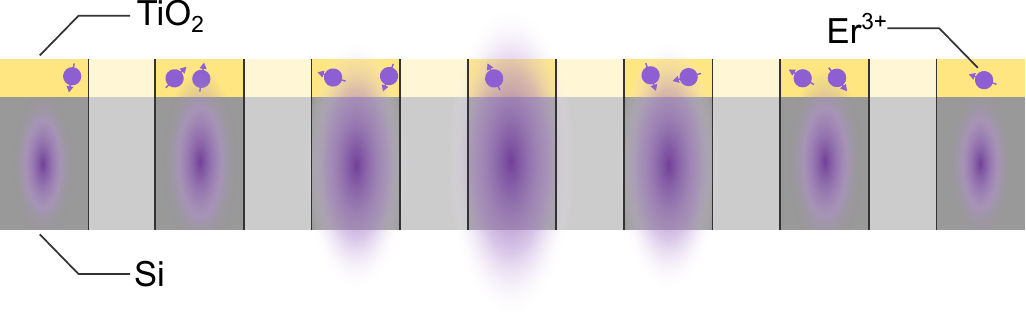}
    \caption{Illustration of a silicon 1D photonic crystal cavity used to enhance the optical emission of erbium ions bound in a TiO$_2$ film grown on top of silicon.}
    \label{fig:purcell}
\end{figure}

In order to optically address single ions within the lowest lying $4f$ - $4f$ transitions, photonic crystal cavities have been utilized to enhance the emission from dilute ensembles of ions as well as resolvable single ions. Enhancement of the optical emission rate has enabled the direct optical observation of single photon emission from the lowest lying optical transitions of rare-earth ions \cite{dibosAtomicSourceSingle2018,zhongOpticallyAddressingSingle2018,yangControllingSingleRare2022,ourariIndistinguishableTelecomBand2023}. The emission enhancement is manifest by a reduction in the excited state lifetime of the ion by increasing the radiative decay rate by a factor $F_P$, known as the Purcell factor. The total excited state decay rate for an ion in a cavity then becomes $\gamma'=F_{P}\gamma_{r} + \gamma_{nr}$. Placing ions in a cavity enhances their emission rate by increasing the density of electromagnetic states available to the ions for optical decay into the cavity compared to ions in the bulk crystal. This heuristic argument leads to a characterization of the Purcell factor by the relationship 

\begin{equation*}
    F_P=\frac{3}{4\pi^2}\frac{\lambda^3}{V}Q,
\end{equation*}

\noindent
where $\lambda$ is the optical wavelength in the crystal medium, $V$ is the mode volume of the cavity mode, and $Q=\nu / \delta\nu$ is the cavity quality factor. The quality factor is characterized by the ratio of the resonant frequency of the cavity mode, $\nu$, to the full width at half maximum of the cavity mode lineshape, $\delta\nu$. From this relationship it becomes clear that cavities with large quality factors and small mode volumes compared to the scale of the optical wavelength help to produce the greatest emission enhancement. In this regard, the silicon photonics platform is particularly appealing because it enables the fabrication of high quality factor, small mode volume cavities that are otherwise unachievable with bulk or free-space optics. Figure \ref{fig:purcell} illustrates a device that couples the evanescent cavity field of a 1D silicon photonic crystal cavity to a thin layer of erbium ions above the cavity surface. State-of-the-art measurements have achieved Purcell factors up to 850 for emission from single Er$^{3+}$ ions \cite{ourariIndistinguishableTelecomBand2023} and radiatively limited single photon emission, where $\gamma'\gg\gamma^*$, has been observed for single Nd$^{3+}$ ions \cite{zhongOpticallyAddressingSingle2018}. Crucially, using Purcell enhancement to increase the optical emission rate of single ions also improves the the indistinguishability of the emitted photons. An indistinguishability of successively emitted single photons up to $I=80\%$ has been observed \cite{ourariIndistinguishableTelecomBand2023}. There is plenty of room for further improvement as well. Photonic crystal cavities with improved design can still significantly increase the achievable Purcell factor though higher cavity quality factors \cite{asanoPhotonicCrystalNanocavity2017} or a reduction of the cavity mode volume \cite{huDesignPhotonicCrystal2016}.

\subsubsection{Challenges}\label{sec:generation_challenges}
One challenge toward realizing a robust source of single photons with rare-earth ion doped materials is the relatively low radiative relaxation rates that lead to low source brightness, as discussed previously in section \ref{sec:considerations_spe} and section \ref{sec:rei_emission}. This particular challenge can be met by placing the ions inside an optical cavity to enhance the radiative decay rate via the Purcell effect to make single ion detection feasible. Progress in working with Purcell enhanced single ions was covered in section \ref{sec:rei_enhancement}. The long lifetimes of the $4f$ states in rare-earth ions make them particularly sensitive to extra dephasing mechanisms within the crystalline host, which requires Purcell factors $F_P > T_2 / T_1$ to realize single photons with a high degree of indistinguishability for quantum networking applications. This challenge can be further compounded by the fabrication of photonic crystal cavities, as the increased proximity to surfaces can introduce spectral wandering and increased optical dephasing. One method to overcome this particular challenge is though passivation of the cavity surfaces \cite{asanoPhotonicCrystalNanocavity2017}. Larger microcavity designs have also been pursued to limit the proximity of ions to the surface \cite{merkelCoherentPurcellEnhancedEmission2020}. Alternatively, these effects can be mitigated by choosing ions with an inversion-symmetric lattice site geometry to reduce the sensitivity of the ion to the fluctuating charges near the surface \cite{wolfowiczQuantumGuidelinesSolidstate2021}.

\subsection{Quantum resource storage}\label{sec:mem}
\subsubsection{Considerations for quantum memories}\label{sec:mem_considerations}
As outlined in section \ref{sec:storage}, different network architectures are possible that utilize quantum memories based on both single ions and ensembles of ions. Each architecture brings certain benefits. For example, memories based on addressable single ions can ultimately be used to construct quantum repeaters that are error corrected \cite{longdellDemonstrationConditionalQuantum2004}, which will be necessary to realize more sophisticated generations of communication networks \cite{muralidharanOptimalArchitecturesLong2016}. On the other hand, memories based on ensembles of ions are particularly useful as they provide a strong interaction with optical photons from the collective action of many ions in the ensemble \cite{duanLongdistanceQuantumCommunication2001}, and are also attractive for multiplexed storage where multiple quantum states can be stored within the same memory \cite{sangouardQuantumRepeatersBased2011,simonQuantumRepeatersPhoton2007,collinsMultiplexedMemoryInsensitiveQuantum2007}. The suitability of a given memory for a particular task can be assessed by the storage time over which the memory can store quantum states, the frequency bandwidth over which the memory can operate, the efficiency with which stored states can be retrieved, and finally the fidelity of the retrieved states to the input state \cite{lvovskyOpticalQuantumMemory2009}. The utility of a quantum memory is particularly sensitive to storage efficiency and the fidelity of the retrieved state. An efficiency above 1/2 and a fidelity greater than 2/3 are required to operate in the no-cloning regime, such that no better copy the retrieved state can exist \cite{grosshansQuantumCloningTeleportation2001,massarOptimalExtractionInformation1995}.

Constructing a memory with an architecture based on rare-earth ions presents several opportunities for storage, both at the single ion level or as an ensemble of ions. Typically, strategies for constructing a single ion memory rely on the spin properties of the ion, while strategies for constructing a memory with an ensemble of ions rely on the optical properties of the ion. In addition to pristine $4f$-$4f$ optical transitions, rare-earth ions possess electronic ground state levels with long spin lifetimes, capable of reaching the order of minutes \cite{hollidaySpectralHoleBurning1993} to days \cite{konzTemperatureConcentrationDependence2003}. In a similar fashion to single photon emission discussed in section \ref{sec:considerations_spe}, the coherence time of the spin is limited by the spin lifetime such that $T_{2} \leq 2T_{1}$. The storage time over which a spin-based memory can store an arbitrary quantum state is directly proportional to the spin $T_2$, so a long spin lifetime is advantageous for the development of quantum memories. Along with their long spin lifetimes, rare-earth ions have demonstrated long spin coherence times \cite{rancicCoherenceTimeSecond2018,ledantecTwentythreeMillisecondElectron2021,ortuSimultaneousCoherenceEnhancement2018,fravalMethodExtendingHyperfine2004}, exceeding 1 second for $^{167}$Er$^{3+}$ in a strong magnetic field \cite{rancicCoherenceTimeSecond2018}. The isotope $^{167}$Er$^{3+}$ is unique among erbium isotopes in that it possesses a nuclear spin $I=7/2$, compared to $I=0$ for all the other stable isotopes of erbium, allowing access to highly coherent hyperfine levels in the presence of a magnetic field. The optical interface provided by the excited states of a rare-earth ion, particularly those within the $4f$ orbital discussed in section \ref{sec:rei_emission}, provide a natural interface to map the state of a photon onto the electronic spin state of an ion. In the following sections we will outline the progress that has been achieved with rare-earth ions in both single ion and ensemble memory architectures. 

\subsubsection{Single rare-earth ion memories}\label{sec:mem_single_rei}
\begin{figure}
    \centering
    \includegraphics{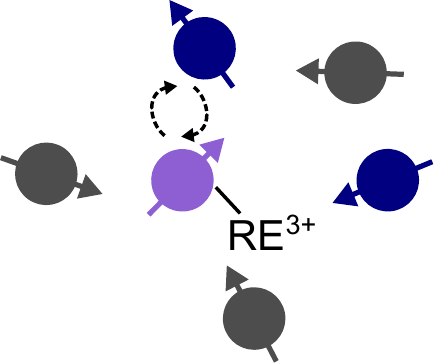}
    \caption{A single rare-earth ion quantum memory. Neighboring nuclear spins (blue) provide an additional resource for long term quantum state storage or even error correction of the electron spin.}
    \label{fig:single_ion_memory}
\end{figure}

Quantum memories based on single defect centers have proved to be a versatile and powerful resource for various quantum systems \cite{awschalomQuantumTechnologiesOptically2018,wolfowiczQuantumGuidelinesSolidstate2021}. A single electron spin can readily serve as a memory for the state of a single photon. Additionally, the state of an electron spin can be mapped onto a nearby nuclear spin for even longer storage \cite{duttQuantumRegisterBased2007,fuchsQuantumMemoryIntrinsic2011} or to enable error correction of the electron spin \cite{taminiauUniversalControlError2014,waldherrQuantumErrorCorrection2014}. Nuclear spins are not naturally sensitive to the optical field of the photon and experience a lower noise environment within the solid state crystal than electron spins, making them attractive candidates to serve as long term memory registers and ancilla qubits for the optically sensitive electron spin qubits. A tradeoff in performance between the different spins must be considered, however, since the much narrower nuclear spin transitions can limit the bandwidth of the memory compared to the wider electron spin transitions. Figure \ref{fig:single_ion_memory} illustrates a single rare-earth ion spin in a crystalline host with possible coupling to adjacent nuclear spins.

Rare-earth ions are suitable for memories based on both electron and nuclear spins. The electron spin states of single ions have been optically controlled \cite{ourariIndistinguishableTelecomBand2023,kindemControlSingleshotReadout2020,siyushevCoherentPropertiesSingle2014}, and spin coherence times for single ions up to 10s of ms have been observed \cite{kindemControlSingleshotReadout2020}. Electron coherence times of this magnitude are already suitable for linking quantum repeater nodes separated by thousands of km \cite{razaviQuantumRepeatersImperfect2009}, with further improvements in coherence time possible with host lattice engineering discussed below in section \ref{sec:mem_challenges}. Additionally, single rare-earth ion electron spins have been coupled to adjacent nuclear spins from the host lattice \cite{kornherSensingIndividualNuclear2020,ruskucNuclearSpinwaveQuantum2022,uysalCoherentControlNuclear2023}. In particular, a single Er$^{3+}$ ion has been used to coherently control a single nuclear spin which served as a register for the electron spin \cite{uysalCoherentControlNuclear2023}, with a SWAP gate fidelity of 87\%. While still early in development, these results indicate that single Er$^{3+}$ ions are a versatile resource for quantum networking, as they can pair C-band telecommunications optical transitions with long spin coherence times from both electron and nuclear spins to serve as repeater memory nodes.

\subsubsection{Ensemble rare-earth ion memories}\label{sec:mem_ensemble_rei}
Quantum memories based on ensembles of rare-earth ions aim to exploit the inhomogeneous broadening of the optical transitions that are naturally present in ion ensembles in the solid-state. In many rare-earth ion doped materials, the inhomogeneous linewidth of the ensemble can be many orders of magnitude larger than the homogeneous linewidth of a single ion \cite{thielRareearthdopedMaterialsApplications2011}. This ratio makes rare-earth ion ensembles particularly well suited for highly precise spectral shaping using spectral hole burning techniques that can enable multimode quantum memories with a large mode capacity \cite{afzeliusMultimodeQuantumMemory2009}. 

The method to turn the inhomogeneous ion ensemble into a multimode memory, as noted in section \ref{sec:storage}, is to write an atomic frequency comb into the ensemble with a modulated narrow band laser. The comb consists of multiple narrow absorption peaks with a frequency separation $\Delta$. If a photon with a spectral bandwidth larger than $\Delta$ and smaller than the width of the comb is incident on the ensemble, it will be absorbed as a collective excitation of the ions as in equation \ref{eqn:AFC_collective}, leading to re-emission at a later time $T=2\pi / \Delta$ in the form of a photon echo \cite{afzeliusMultimodeQuantumMemory2009,deriedmattenSolidstateLightMatter2008}. This basic functionality forms the backbone of atomic frequency comb memories. While the memory described so far provides only a fixed-duration storage before recall at time $T$, the echo photon can be recalled on-demand by using an optical control pulse to transfer the collectively excited ions into an auxiliary ground state spin level for storage as a spin wave. Recall of the echo photon from the spin wave can then be made with the application of another control pulse \cite{afzeliusMultimodeQuantumMemory2009,afzeliusDemonstrationAtomicFrequency2010}.

Atomic frequency comb memories have been used to great effect with rare-earth ion ensembles. Multimode memories have been realized that preserve quantum correlations between highly multiplexed photons stored in the near infrared \cite{busingerNonclassicalCorrelations12502022} and telecom \cite{weiQuantumStorage16502023} bands, where Er$^{3+}$ ions provide storage in the telecom band. Additionally, fidelities of the recalled state to the input state of up to 99.3\% have been realized \cite{zhouQuantumStorageThreeDimensional2015}, along with qubit storage times up to 10s of ms \cite{ortuStoragePhotonicTimebin2022}. Efficiencies of 53\% \cite{jobezCavityenhancedStorageOptical2014} and 56\% \cite{sabooniEfficientQuantumMemory2013} have been realized for classical input pulses, while efficiencies for quantum light storage up to 27.5\% \cite{davidsonImprovedLightmatterInteraction2020} have been realized, all without on-demand recall. Memory efficiency with on-demand recall using spin wave storage has been demonstrated with an efficiency up to 12\% \cite{jobezCavityenhancedStorageOptical2014}.

Atomic frequency comb memories are also being explored in nanophotonic platforms \cite{craiciuNanophotonicQuantumStorage2019,zhongNanophotonicRareearthQuantum2017}, critically demonstrating that the protocol is robust even as the physical size of the memory shrinks. The ability to scale the size of quantum memories down from bulk crystals to nanofabricated devices, and especially to devices that can be fabricated with standard semiconductor foundry processes, will be a necessary step to realize repeater networks that span a quantum internet. Improvements to atomic frequency comb memory performance are progressing rapidly and demonstrate the potential for rare-earth ion ensembles to deploy as multimode memories in quantum networks. Here we note that other protocols for building quantum memories with ensembles of ions, such as gradient echo memories \cite{hedgesEfficientQuantumMemory2010}, electromagnetically induced transparency \cite{longdellStoppedLightStorage2005}, and off resonant Raman techniques \cite{nunnMappingBroadbandSinglephoton2007} are also being developed, but their multimode capacity can be severely limited by the small size and optical depth of nanophotonic devices. Conversely, the multimode capacity of atomic frequency comb memories is independent of optical depth \cite{afzeliusMultimodeQuantumMemory2009}, making the atomic frequency comb protocol appealing for use in state-of-the-art nanophotonic devices. 

\subsubsection{Challenges}\label{sec:mem_challenges}
There are still several outstanding challenges toward realizing the goal of widely deployed quantum memories based on rare-earth ion technologies. In particular, the choice of host lattice has a large effect on the spin coherence of the memory, both for single and ensembles of ions. Despite ion ensembles demonstrating memory performance through the atomic frequency comb protocol, which provides storage in the optical transitions, the need for deterministic recall of the echo photons in repeater networks necessitates the use of spin wave storage. The storage time of the spin wave is dictated by the electron spin coherence of the ion. The performance of quantum memories for both single ion and ensemble architectures therefore depend critically on the achievable spin coherence of the ion. 

In order to obtain long spin coherence times, such that the spin coherence can approach the lifetime limit, it is advantageous to minimize the interaction of the ion with its host lattice. One route toward achieving this is through the use of non-Kramers ions, which have an even number of electrons with a quenched electronic angular momentum. These ions, including Pr$^{3+}$ and Eu$^{3+}$, have demonstrated the required long lifetimes and coherence times for use as quantum memories \cite{hollidaySpectralHoleBurning1993,konzTemperatureConcentrationDependence2003,fravalMethodExtendingHyperfine2004,longdellStoppedLightStorage2005,hedgesEfficientQuantumMemory2010,jobezCoherentSpinControl2015,gundoganSolidStateSpinWave2015}, but none of these ions have optical transitions compatible with the telecom transmission band. Kramers ions, such as Nd$^{3+}$, Er$^{3+}$, and Yb$^{3+}$, with an odd number of electrons, are more susceptible to their electromagnetic environment and typically demonstrate shorter spin lifetimes and coherence times as a result. However, Er$^{3+}$ provides critical optical transitions within the telecom transmission band, making it essential to demonstrate long spin coherence times in spite of the significant electron spin interactions with the host lattice. This challenge demonstrates the need for developing host lattices that can minimize the impact to the electron spin.

The use of host lattices with low densities of paramagnetic impurities and nuclear spins are particularly effective for extending spin coherence \cite{kanaiGeneralizedScalingSpin2022}. Many common host lattices for rare-earth ions, such as yttrium orthosilicate (Y$_2$SiO$_5$; YSO), yttrium orthovanadate (YVO$_4$; YVO), and yttrium aluminium garnet (Y$_3$Al$_5$O$_{12}$; YAG), often contain residual concentrations of other rare-earth ion dopants and impurities in addition to the desired species and also have a fixed background of yttrium nuclear spins ($I=1/2$) that cannot be isotopically purified. While impressive demonstrations of spin coherence have been made in these materials, many have required the use of dynamical decoupling sequences or challenging magnetic field configurations which add extra overhead to the memory and limits scalability. This problem is being addressed by a search for materials that can host rare-earth ions with limited or no background spins \cite{stevensonErbiumimplantedMaterialsQuantum2022,ferrentiIdentifyingCandidateHosts2020}, such as titanium dioxide (TiO$_2$) \cite{dibosPurcellEnhancementErbium2022,phenicieNarrowOpticalLine2019} and calcium tungstate (CaWO$_4$) \cite{ourariIndistinguishableTelecomBand2023,ledantecTwentythreeMillisecondElectron2021} among others. Nanoparticles and molecular crystals are also being considered as potential hosts for rare earth ions \cite{alqedraOpticalCoherenceProperties2023,serranoUltranarrowOpticalLinewidths2022,zhongEmergingRareearthDoped2019}. As the search for new host crystals progresses, it is important to factor in compatibility with semiconductor foundry processes, so that advances in host selection can leverage the scalable production of semiconductor manufacturing.

An additional challenge for all rare-earth ion based memory architectures is achieving suitable coupling between common single photon sources, such as spontaneous parametric down conversion and four wave mixing sources or semiconductor quantum dots, and the rare-earth device. These sources often have a large bandwidth that contrasts with the typically narrow bandwidth of rare-earth ion systems, leading to reduced efficiency in the storage process. While efficiency can be improved with impedance-matched cavity designs \cite{davidsonImprovedLightmatterInteraction2020,zhongNanophotonicRareearthQuantum2017,jobezCavityenhancedStorageOptical2014,sabooniEfficientQuantumMemory2013,afzeliusImpedancematchedCavityQuantum2010}, the challenge of bandwidth can be addressed in ensemble memories with larger inhomogeneous broadening, particularly by using frequency combs of larger bandwidth \cite{davidsonImprovedLightmatterInteraction2020}. Conversely, these challenges can potentially be met by proper tailoring of the input photons to be stored, or by storing photons produced by rare-earth ion sources \cite{dibosAtomicSourceSingle2018,zhongOpticallyAddressingSingle2018,yangControllingSingleRare2022,ourariIndistinguishableTelecomBand2023}.

\subsection{Quantum resource distribution}\label{sec:dist}
\subsubsection{Considerations for quantum resource distribution and network construction}\label{sec:dist_considerations}
Quantum resource distribution is the keystone with which quantum networks are ultimately built. The fundamental operation that enables quantum resource distribution is the performance of entanglement swapping operations that create links between remote nodes which were discussed in section \ref{sec:distribution}. Quantum resource distribution relies directly on the performance of the repeater as both a source (see sections \ref{sec:ent_generation} and \ref{sec:resource_generation}) and memory (see sections \ref{sec:storage} and \ref{sec:mem}), as well as external factors involved with overall network design such as node spacing and the degree of node connectivity. Given the interplay between repeater performance metrics and achievable network design, it is useful to adopt a co-design perspective where the needs of the network at the system level can be used to inform design choices at the device level and vice versa \cite{singhKeyDeviceMaterials2021}. 

Despite recent advances in demonstrating a multi-node quantum network enabled by atom-like solid-state systems \cite{pompiliRealizationMultinodeQuantum2021}, it remains an outstanding challenge to demonstrate entanglement distribution rates \textit{with} quantum repeaters that exceed the rates possible \textit{without} the use of quantum repeaters \cite{awschalomRoadmapQuantumInterconnects2022}. Meeting this challenge will require significant development in the state-of-the-art for quantum repeater performance, as well as in semiconductor manufacturing capabilities. Increasing the scalability of solid-state quantum platforms to the necessary level, including rare-earth ion based platforms, may require semiconductor foundries that can work with new materials and new process design kits dedicated to quantum device design. In addition to the integration of quantum sources, memories, control and measurement devices in a unified, low-loss platform, it will also be necessary to multiplex the components to achieve useful entanglement rates \cite{collinsMultiplexedMemoryInsensitiveQuantum2007,simonQuantumRepeatersPhoton2007,munroQuantumMultiplexingHighperformance2010}. These challenges underscore the need for scalable design at the device level informed by network requirements. 

Ultimately the success of quantum networks, and rare-earth ion platform for quantum networks in particular, will hinge upon whether or not components and devices can be made reliably at the scale necessary to support the networks. Progress with the fabrication of rare-earth ion devices in platforms compatible with large-scale semiconductor foundry processes suggests that this is possible \cite{przybylinskaOpticallyActiveErbium1996,kenyonErbiumSilicon2005,weissErbiumDopantsNanophotonic2021,gritschNarrowOpticalTransitions2022,berkmanObservingEr32023,staudtInterferenceMultimodePhoton2007,sinclairSpectroscopicInvestigationsTi2010,saglamyurekBroadbandWaveguideQuantum2011,duttaIntegratedPhotonicPlatform2020,yangControllingSingleRare2022,gongLinewidthNarrowingPurcell2010,gongObservationTransparencyErbiumdoped2010,dingMultidimensionalPurcellEffect2016,singhEpitaxialErdopedSilicon2020,dibosPurcellEnhancementErbium2022}. An important avenue for development to improve achievable entanglement distribution rates is with the development of repeaters based on single rare-earth ions that have the potential to improve rates with increased entanglement swapping probability \cite{asadiProtocolsLongdistanceQuantum2020, asadiQuantumRepeatersIndividual2018}. Experiments demonstrating single shot readout of individual ions \cite{kindemControlSingleshotReadout2020,rahaOpticalQuantumNondemolition2020}, along with multiplexing of individual ions in a single device \cite{chenParallelSingleshotMeasurement2020,ulanowskiSpectralMultiplexingTelecom2022} suggest that there is a path forward for truly scalable rare-earth ion repeaters.  

\subsubsection{Near-term benchmarks for rare-earth ion enabled quantum networks}\label{sec:dist_benchmarks}
In order to support and guide the development of rare-earth ion repeaters for quantum networking, it is useful to determine a set of benchmarks for near-term performance. These benchmarks should be guided by the current state-of-the-art with knowledge of existing material limitations and network capabilities and will help inform benchmarks for longer term development. Here we limit these benchmarks to foundry compatible platforms using single erbium ions, as these are the best candidates for large scale fabrication and interfacing with low-loss optical fiber. 

The ultimate goal of constructing a global quantum internet connected by quantum repeaters will likely require the realization of smaller scale networks first. These networks may consist of modular interconnects that connect nodes within a single room, to data center networks or or local area networks that connect nodes up to 100s of meters \cite{awschalomRoadmapQuantumInterconnects2022,awschalomDevelopmentQuantumInterconnects2021}. Addressing networks of this size will be an important step in developing the technologies needed to span more global networks. For these distances, optical and spin coherence times on the order of 10~$\mu$s are suitable for coherent interactions between remote ions and achieving these benchmarks in scalable, nanofabricated devices suitable for the desired networking tasks becomes the challenge. Particularly, developing nanofabricated devices with component efficiencies capable of operating in the no-cloning regime \cite{grosshansQuantumCloningTeleportation2001,massarOptimalExtractionInformation1995} will be critical. 

Here we note that optical coherence times up to 10.2~$\mu$s have been observed with a Hahn echo technique for single erbium ions in CaWO$_4$ coupled to silicon nanocavities, and the same platform has demonstrated electron spin coherence up to 44~$\mu$s also using a Hahn echo technique, likely limited by other paramagnetic impurities in the host crystal \cite{ourariIndistinguishableTelecomBand2023}. Single cavity coupled erbium ions in Y$_2$SiO$_5$ have also been coupled with single neighboring nuclear spins with coherence times up to 1.9 ms \cite{uysalCoherentControlNuclear2023}. These results demonstrate that erbium ions coupled to nanophotonic structures are capable of reaching the necessary coherence times for local network applications. The primary roadblock to scalability is the bulk crystal used to house the ions. Demonstrations of optical and spin coherence times up to 10~$\mu$s in a platform that eliminates the need for the bulk crystal would enable rare-earth ion technologies to begin addressing the needs of networks at this scale. Developing global scale quantum networks will require broad based advances in both hardware, software, and network design structures that will continue to happen over a longer time period (likely exceeding 10 years) \cite{awschalomRoadmapQuantumInterconnects2022,awschalomLongdistanceEntanglementBuilding2020}. We expect that advances in scalable rare-earth ion platforms today will have significant impact on the development of these larger networks.

\section{Perspectives}\label{sec:perspectives}
The development of a fully scalable quantum repeater will enable the growth of quantum networks with nodes spanning 100s of km or more, which will form the backbone of a quantum internet. Quantum repeaters require the ability to faithfully generate, store, and distribute entanglement between the network nodes, and rare-earth ions embedded in solid-state hosts, particularly hosts compatible with large scale semiconductor manufacturing, are an attractive platform to meet these requirements. We have drawn special attention to erbium in this review because it provides a natural interface with low-loss optical telecommunications networks operating near 1550 nm. Of particular near-term importance will be the development of erbium ion devices in a scalable platform that can achieve optical and spin coherence times on the order of 10~$\mu$s. This will enable quantum networks at the local level with nodes that span up to 100s of meters, spurring continued innovation in quantum networking technologies. The continued development of rare-earth ion technologies in scalable platforms has the capability to bring budding quantum networks to fruition.

Taking a broader view, the large ecosystem of quantum technologies presents additional challenges and opportunities. Quantum computers and quantum sensors are all being developed with different and often incompatible platforms, from superconducting and spin qubits operating at mK temperatures and GHz frequencies, to trapped ion and atom systems operating in ultra-high vacuum and frequencies of 100s of THz. In order to build robust and versatile networks capable of harnessing the best qualities of each system, it will be necessary to establish ways to form interconnects between them \cite{awschalomDevelopmentQuantumInterconnects2021,awschalomRoadmapQuantumInterconnects2022}. 

A particular interconnect, the quantum state transducer, would be capable of mapping the quantum state of a qubit in one platform onto a qubit in a separate platform and vice versa. The importance of connecting quantum devices to fiber networks for long range networking means one of these platforms is often required to be an optical photon with a wavelength near 1550 nm to access the low-loss telecom transmission window. A solution for quantum state transduction between the microwave and telecom optical regimes \cite{laukPerspectivesQuantumTransduction2020} is critically needed for superconducting \cite{kjaergaardSuperconductingQubitsCurrent2020} and spin \cite{vinetPathScalableQuantum2021} qubit platforms for quantum computing. 

Rare-earth ion systems may play a significant role in this regard since they possess highly coherent microwave and optical frequency transitions. Schemes to realize microwave to optical transducers with high efficiency using rare-earth ions have been proposed \cite{williamsonMagnetoOpticModulatorUnit2014,kimiaeeasadiProposalTransductionMicrowave2022} and experiments are actively pursuing the goal \cite{rochmanMicrowavetoopticalTransductionErbium2023,bartholomewOnchipCoherentMicrowavetooptical2020,fernandez-gonzalvoCavityenhancedRamanHeterodyne2019,welinskiElectronSpinCoherence2019}. The promise of quantum state transducers using rare-earth ions indicates that the rare-earth ion platform is not only attractive for developing the quantum repeaters necessary to build out global quantum networks, but also for providing a means to expand the reach of the network to other diverse qubit platforms and technologies. This expanded toolkit will ultimately lead to more functional quantum networks by allowing qubits of different types to interact within the same network. Rare-earth ions are therefore well positioned to play a significant role in a future quantum internet.

\section*{Acknowledgments}
Work by S.E.S. and M.K.S. was carried out at Argonne National Laboratory with support from the US Department of Energy Office of Science Advanced Scientific Computing Research program under CRADA A22112 through the Chain Reaction Innovations program. The authors would like to further acknowledge helpful conversations with Christophe Jurczak and Xavier Aubry.

\bibliographystyle{quantum}
\bibliography{REIperspective}

\end{document}